# Improving antibody language models with native pairing

Sarah M. Burbach[1,2,3], Bryan Briney[1,2,3,4,5]

[1]Department of Immunology and Microbiology, The Scripps Research Institute, La Jolla, CA 92037 USA
[2]Center for Viral Systems Biology, The Scripps Research Institute, La Jolla, CA 92037 USA
[3]Multi-Omics Vaccine Evaluation Consortium, The Scripps Research Institute, La Jolla, CA 92037 USA
[4]Scripps Consortium for HIV/AIDS Vaccine Development, The Scripps Research Institute, La Jolla, CA 92037 USA
[5]San Diego Center for AIDS Research, The Scripps Research Institute, La Jolla, CA 92037 USA

Correspondence may be directed to: briney@scripps.edu

**BIGGER PICTURE**
Antibodies are broadly used as therapeutics against a variety of human diseases. Antibody discovery, evaluation, and clinical development would be greatly accelerated by computational models able to accurately infer structural and functional characteristics of antibodies directly from their sequence. One particularly promising approach is the training of antibody-specific large language models, which can serve as the foundation for fine-tuning specialized predictive models. Human antibodies comprise a unique pairing of a single heavy chain and light chain, with both chains contributing to the antigen binding region, or paratope, of the antibody. Current state-of-the-art antibody language models have been trained exclusively using unpaired sequence data, however, and are unable to learn cross-chain features that are necessary to fully understand structure and function. By training a set of models using paired or unpaired antibody sequence data, we show that training with natively paired antibody sequences improves model performance across a variety of metrics and tasks, including pathogen specificity classification. We also demonstrate a potential strategy for overcoming the paucity of natively paired antibody sequence datasets by fine-tuning the general protein language model ESM-2 and showing that the fine-tuned model learns immunologically relevant, cross-chain antibody features.

**SUMMARY**
Current antibody language models are limited by their use of unpaired antibody sequence data and the biases in publicly available antibody sequence datasets, which are skewed toward antibodies against a relatively small number of pathogens. A recently published dataset (by Jaffe, et al) of approximately $1.6 \times 10^6$ natively paired human antibody sequences from healthy donors represents by far the largest dataset of its kind. It offers a unique opportunity to evaluate how antibody language models can be improved by training with natively paired antibody sequence data. We trained two Baseline Antibody Language Models (**BALM**), using natively paired (**BALM-paired**) or unpaired (**BALM-unpaired**) sequences from the Jaffe dataset. We provide evidence that training with natively paired sequences substantially improves model performance on a variety of metrics and that this improvement results from the model learning immunologically relevant features that span the light and heavy chains. We fine-tuned ESM-2, a state-of-the-art general protein language model, with natively paired antibody sequences (**ft-ESM**) and showed that it learns similar cross-chain features as BALM-paired. Finally, we evaluated the ability of each model to classify antibodies by pathogen specificity and demonstrate that training with natively paired antibody sequences substantially improves downstream task performance.

**INTRODUCTION**

It is estimated that the circulating antibody repertoire is composed of as many as $10^{18}$ unique Abs [1], which surpasses the combined number of unique proteins encoded by all the genomes of all species on earth by many orders of magnitude [2]. The extraordinary diversity of the human antibody repertoire is produced initially by somatic recombination of germline gene segments [3]. Antibody heavy chains are assembled from variable (**V**), diversity (**D**), and joining (**J**) gene segments. Light chains are assembled similarly, but without D gene segments. This recombination process occurs independently in each B cell and the resulting antibody is expressed as a dimer of heterodimers, containing two identical heavy chains and two identical light chains. The antigen binding regions of the antibody, which determine antigen specificity, are each composed of six complementary determining region (**CDR**) loops: three encoded by the heavy chain and three by the light chain.

Further diversification of antibodies occurs upon exposure to a non-self antigen, when B cells encoding antigen-specific antibodies undergo an iterative affinity maturation process that consists of multiple rounds of clonal expansion, somatic hypermutation (**SHM**), and antigen driven selection [4–6]. Through this process, antigenic stimulation of a single naive B cell can produce a *clonal lineage* of B cells, each expressing an antibody that is related to the parental antibody but which has accumulated a unique set of somatic mutations. These affinity matured antibodies often contain only a handful of deviations from the original germline recombination, but affinity is typically improved by several orders of magnitude [7]. Following antigen clearance, a subset of B cells encoding affinity matured, antigen-specific antibodies are retained as an *immune memory* of the encounter [8,9], which allows rapid response to subsequent exposure and is the primary mechanism of protection for most vaccines. In essence, each person's unique collection of affinity matured antibody genes constitutes a detailed molecular record of all previous pathogen encounters.

Much as the meaning of a sentence is determined by the order and context of its words, the structure and function of a protein is encoded by its amino acid sequence. More concisely, ***sequence* determines *structure* determines *function*** [10]. The conceptual similarity between language and biological sequences inspired the application of language models (**LMs**) to biological sequence data, with the goal of gaining a deeper understanding of the language of proteins [11]. LMs trained on general protein sequence data (**PLMs**), such as HelixFold and ESMFold, have successfully learned information about evolutionary fitness, function, and structure [12–14]. This suggests the models have learned a deep understanding of the fundamental properties of amino acids and the importance of the order and context in which they occur. Applying PLMs to antibody sequences yielded some success, but PLMs generally exhibited only a cursory understanding of antibodies that did not extend beyond "obvious" features such as germline gene use [15,16].

Antibody-specific LMs (**AbLMs**), which use essentially unmodified LM or PLM model architectures but are trained using antibody sequence data, have learned features like SHM [16–18] and are substantially better than PLMs at antibody sequence infilling [16] These results indicate that AbLMs possess a more sophisticated understanding of features that differentiate antibodies from the general protein space and provide a strong argument for training specialized AbLMs instead of repurposing pre-trained PLMs. However, AbLMs still have vast room for improvement.

ESMFold and HelixFold demonstrate that existing model architectures can support powerful biological LMs. Thus, the primary factors impeding AbLM development are instead related to the lack of suitable training data at a sufficient scale. First, all existing transformer-based

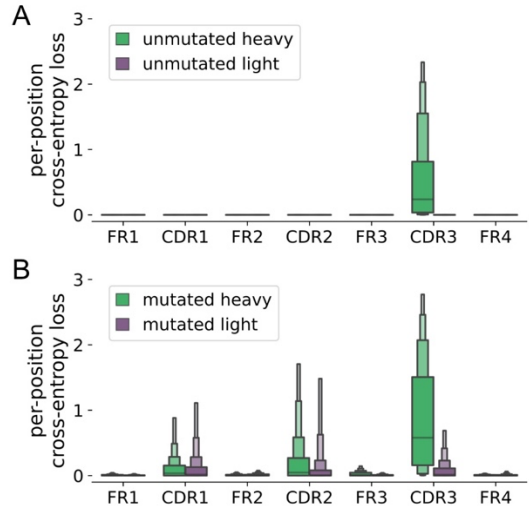

**Figure 1. Per-position cross-entropy loss of BALM-paired.** Per-position cross-entropy loss was calculated by iteratively masking each position and predicting the masked residue with BALM-paired using either unmutated (A) or mutated (B) test sequences. For each sequence, the median cross-entropy loss was computed for each FR or CDR, and the distribution of median values is shown using a letter value plot [24].

AbLMs were trained using only unpaired antibody sequences. This was by necessity rather than design; the far lower cost of generating unpaired sequences means there are orders of magnitude more unpaired than natively paired antibody sequences available [19,20]. Nevertheless, AbLMs trained using only unpaired data cannot learn cross-chain features that encode important information about antibody structure and function. Second, publicly available antibody datasets are skewed toward a relatively small number of disease states, including autoimmunity, cancer, and infectious diseases like HIV, influenza, and SARS-CoV-2. This produces AbLMs with a parochial view of the Ab repertoire rather than a complete understanding of Ab diversity.

A recently published dataset of approximately 1.6 million natively paired human antibody sequences [21,22] provides an opportunity to assess the value of training an AbLM with natively paired data. This unique dataset from Jaffe *et al*, which is the largest publicly available collection of natively paired human antibody sequences, was compiled using circulating B cells from healthy adult donors to produce a minimally skewed representation of the baseline human antibody repertoire. The Jaffe dataset is much smaller than the unpaired datasets used to train existing AbLMs, however, and it is unlikely that the training advantages of native pairing are sufficient to overcome this massive difference in scale. Thus, the goal of this work is to determine whether and to what extent an AbLM can be improved by training with natively paired antibody sequence data rather than unpaired sequence data.

To accomplish this, we trained two baseline antibody language model (**BALM**) variants using training datasets that were identical except for their inclusion (**BALM-paired**) or exclusion (**BALM-unpaired**) of antibody pairing information. Although both models rapidly learn features associated with recombination and maturation, BALM-paired performs substantially better than BALM-unpaired across a variety of metrics, with notable improvements in the information content of light chain embeddings. We further demonstrate that the improved performance of BALM-paired is linked to its ability to learn features that span the heavy and light antibody chains. We additionally fine-tuned an ESM-2 model (**ft-ESM**) with the same paired sequences, to demonstrate a potential middle-ground approach given the limited nature of paired data.

Finally, we show that the paired models, BALM-paired and ft-ESM, exhibit improved performance on three specificity classification tasks.

## RESULTS

**Training a baseline antibody language model.** BALM-paired and BALM-unpaired use a slightly modified RoBERTa-large architecture and were trained with a masked language model objective on 1,335,854 antibody sequence pairs [22]. BALM-paired was trained on natively paired sequences, with concatenated heavy and light chains split using a separator token. BALM-unpaired was trained on light and heavy chains separately, with only one chain per input. To equalize training steps between the models, BALM-unpaired was trained using a batch size of 512 which is twice that of BALM-paired, at 256.

**BALM rapidly learns germline antibody features.** The combinatorial diversity of antibody recombinants (that is, the diversity provided by the selection of individual V, D and J genes for recombination) is relatively small compared to the diversity contributed by non-templated addition and deletion at recombination junctions [23]. Thus, it is expected that AbLMs will learn germline-encoded features more readily than the more complex patterns inherent in non-templated regions. To assess this on BALM-paired, we separately analyzed the per-position cross-entropy loss (**CEL**) of mutated or unmutated sequences (*Fig 1*). By grouping sequence positions into their corresponding framework region (**FR**) or complementarity determining region (**CDR**), we observed much weaker model performance in the untemplated CDR3s of unmutated sequences. Additionally, we observed moderately lower model performance in all regions of mutated antibody sequences, which contain untemplated somatic mutations distributed throughout the sequence. Antibody mutations are clustered in CDRs, and BALM-paired performs substantially less well in the CDRs of mutated sequences compared to the relatively less mutated FRs.

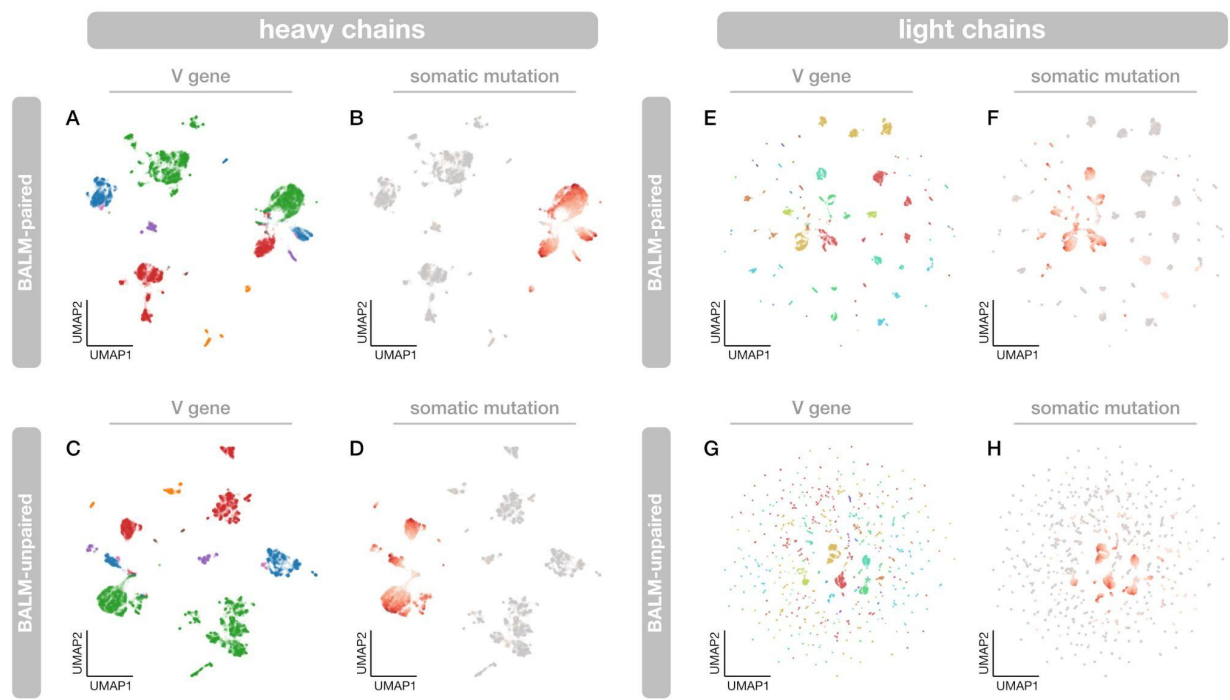

**Figure 2. Training with natively paired sequence data improves light chain embeddings**. UMAP of final layer embeddings for heavy chains (A-D) and light chains (E-H), colored by V-gene or number of somatic mutations, for BALM-paired (top row of plots) and BALM-unpaired (bottom row of plots).

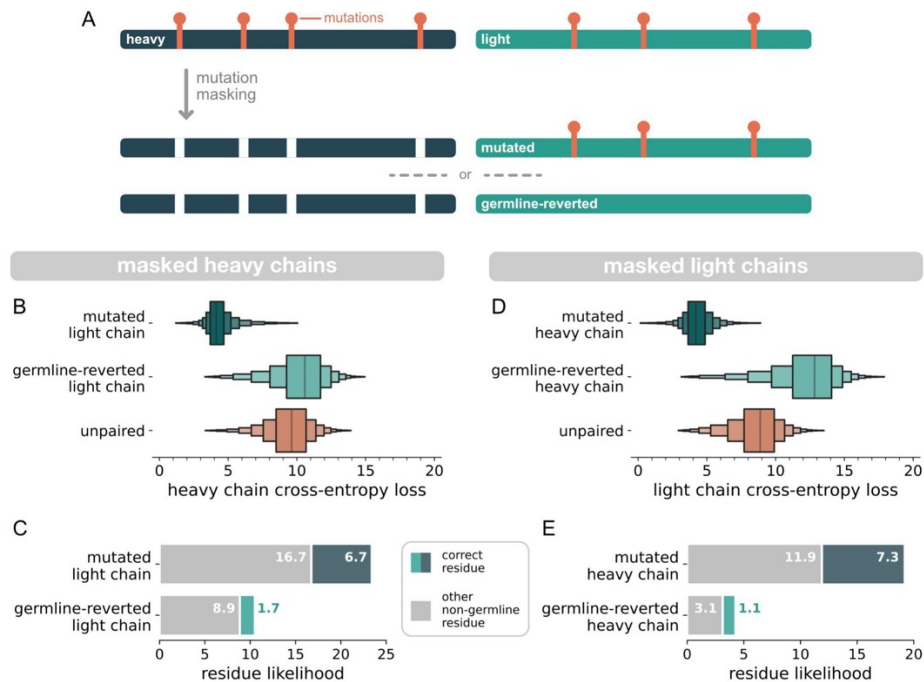

**Figure 3. Cross-entropy loss of masked mutations in light and heavy chains.** (A) Schematic of the mutation masking process, in which the mutated positions in a single chain are masked and the masked chain is paired with either the native (mutated) partner chain or a germline-reverted variant of the partner chain. (B) Cross-entropy loss of masked mutations in heavy chains, when paired with mutated light chain and germline reverted light chain for BALM-paired, and alone for BALM-unpaired. (C) Likelihood (model output probabilities, multiplied by 100) assigned by BALM-paired to the correct masked heavy chain residue (blue: mutated light chain, green: germline-reverted light chain) or to any other non-germline residue (gray). Likelihood values are the average of all masked positions across all test sequences. (D) Cross-entropy loss of masked mutations in light chains, when paired with mutated heavy chain and germline reverted heavy chain for BALM-paired, and alone for BALM-unpaired. (E) Likelihood assigned by BALM-paired to the correct masked light chain residue (blue: mutated heavy chain, green: germline-reverted heavy chain) or to any other non-germline residue (gray). Likelihood values are the average of all masked positions across all test sequences.

**Native pairing preferentially improves light chain embeddings.** The previously reported AbLMs AntiBERTa [18] and AbLang [16] have shown that clustering the output embeddings of these models can group antibody sequences according to V gene use and SHM. Despite both of these AbLMs being trained on datasets that include light chains, only heavy chain embeddings were analyzed. Using a test dataset of 20,000 natively paired antibody sequences, we analyzed the output embeddings of BALM-paired and BALM-unpaired. As described previously [16,18], embeddings from the final transformer layer were averaged along the input length dimension and a uniform manifold approximation and projection (**UMAP**) representation was computed [25,26]. Because BALM-paired output embeddings include both heavy and light chains, we extracted a subset of the BALM-paired embedding which contains only the positions corresponding to a single chain (either heavy or light) before averaging over the length dimension. This allows us to directly compare the embeddings produced by BALM-paired and BALM-unpaired.

Heavy chain embeddings from BALM-paired and BALM-unpaired clustered similarly, grouping sequences primarily by mutation and secondarily by V gene (***Fig 2a-d***). This mirrors results seen with AntiBERTa [18] but differs slightly from AbLang [16], for which output embeddings cluster primarily by V gene and secondarily by mutation. Clustered light chain embeddings of the two models were quite different, however (***Fig 2e-h***). Although BALM-unpaired embeddings of mutated light chain sequences form reasonably well-defined V gene clusters (***Fig 2g***), unmutated light chain embeddings were essentially randomly dispersed (***Fig 2h***). In contrast, the clustered light chain embeddings produced by BALM-paired are similar to heavy chains, segregating sequences primarily by mutation and secondarily by V gene (***Fig 2e, 2f***). This suggests that BALM-paired is learning cross-chain features and that these features preferentially improve model performance with antibody light chains.

**Paired model improvements are driven by learning cross-chain features.** We next sought to more deeply investigate BALM-paired's ability to learn features that span both antibody chains. From our held-out test dataset, we selected all sequence pairs containing at least 3 mutations in each of the heavy and light chains. All mutated heavy chain positions were masked and BALM-paired was asked to predict the masked residues when the heavy chain was paired with (***1***) the natively paired light chain, or (***2***) a germline-reverted version of the light chain in which all mutated light chain residues were reverted to germline. For comparison, BALM-unpaired was also asked to predict the same masked residues given only the unpaired heavy chain sequence. We noted a large reduction in CEL when the masked heavy chain was paired to the native (mutated) light chain (***Fig 3a***), indicating that native pairing improved model performance through cross-chain learning. BALM-paired considered the correct (mutated) residue approximately 4-fold likely when the masked heavy chain was paired with the native light chain (6.7 vs 1.7), and also considered incorrect but non-germline residues about twice as likely when the masked heavy chain was paired with the native light chain (16.7 vs 8.9), indicating that the model is learning patterns of somatic mutation rather than memorizing specific mutations (***Fig 3b***). Results from the reciprocal experiment, in which light chains were masked and paired with native or germline-reverted heavy chains (***Fig 3c-d***), were even more striking: native pairing increased the likelihood of the correct (mutated) residue by over 6-fold (7.3 vs 1.1) and the likelihood of any non-germline residue by nearly 5-fold (11.9 vs 3.1).

**Learned cross-chain features are immunologically relevant.** The extremely high cost of generating natively paired antibody datasets led us to evaluate whether a general protein LM could learn similar cross-chain features by fine-tuning with natively paired antibody sequences. The pre-trained 650-million parameter ESM-2 model [14] was fine-tuned with a MLM objective on the same dataset of 1,335,854 paired antibody sequences used to train BALM-paired [22]. After fine-tuning, we evaluated several clinically approved therapeutic monoclonal antibodies (**mAbs**) using both the base ESM-2 model and our fine-tuned ESM-2 variant. Separately for each mAb, we averaged the attention values for each pair of heavy/light positions across all attention heads of all model layers to produce a single cross-chain attention matrix. Results from the representative mAb Masavibart are shown in ***Figure 4***, and data for several additional mAbs can be found in ***Figure S1***. The fine-tuned ESM-2 (**ft-ESM**) model focuses its attention on CDRs (***Fig 4a***), with particular attention directed toward the heavy chain CDR3 and conserved Cys residues in both antibody chains (***Fig 4e***). This corresponds to immunologically relevant regions of the antibody sequence responsible for antigen binding, as well as structural regions where the heavy and light chains are in close proximity (***Fig 4b***). In contrast, the base ESM-2 model directs its cross-chain attention on residues that are proximal in the linear input sequence rather than structurally or immunologically relevant residues (***Fig 4c***). Since model inputs were concatenated heavy and light chains split by a separator token, this corresponds to heightened attention on residues near the end of the heavy chain and the start of the light chain (*Fig 4f* and lower left corner of *Fig 4c*). Unsurprisingly for a general protein LM, base ESM-2 also pays substantial attention to cysteine residues. Because the base ESM-2 model does not preferentially attend to immunologically relevant positions prior to fine-tuning, the cross-attention patterns being learned must be a direct result of fine-tuning with natively paired sequences.

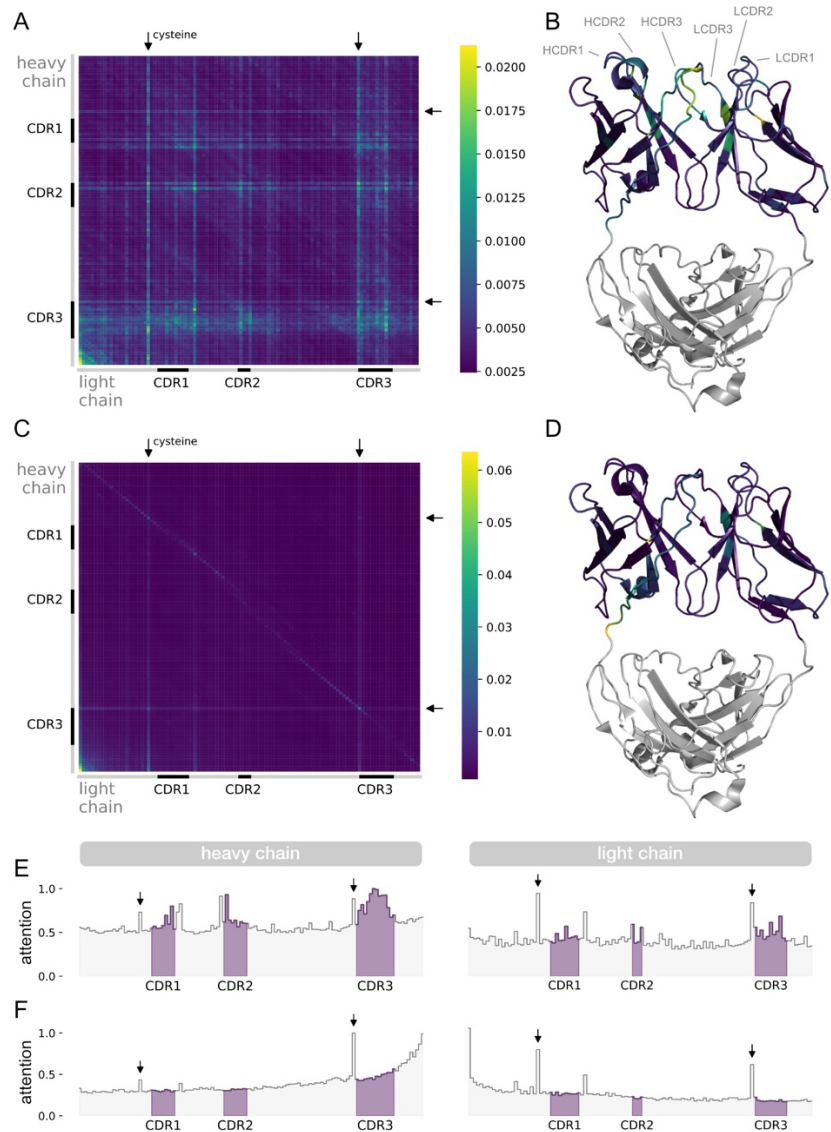

**Figure 4. Cross-chain attention of ESM-2 before and after fine-tuning with paired antibody sequences.** Cross-chain attention matrices were computed for the clinically-approved anti-SARS-CoV-2 mAb Masavibart by averaging cross-chain attention across all model layers and attention heads, using either (A) the fine-tuned ESM-2 model or (C) the base ESM-2 model. Mapping per-position cross-chain attention of the fine-tuned ESM-2 model onto the Masavibart structure (PDB: 6xdg) revealed a focus on structural regions important for antigen recognition (B and E). In contrast, per-position cross-chain attention of the base ESM-2 model was focused primarily on cysteine residues and on positions near the end of the heavy chain or the start of the light chain, which are proximal in the concatenated input sequence and distant from the antigen recognition site (C and F). To demonstrate that the results observed for Masavibart are representative, cross-chain attention matrices for four additional anti-SARS-CoV-2 mAbs can be found in Supplementary Figure 1.

**Training with natively paired antibody sequences improves specificity classification.**
BALM-unpaired, BALM-paired, ft-ESM and the 650M-parameter ESM-2 (**base-ESM**) were fine-tuned with a sequence classification head to perform three separate antibody specificity classification tasks. The first task, trained on approximately 20,000 paired antibodies (~10,000 in each class), was a binary classification of coronavirus (**CoV**)-specific antibodies against a collection of randomly selected antibodies from the memory B cell repertoires of several healthy donors. ft-ESM was the best performer across all metrics, followed closely by BALM-paired which outperforms both base-ESM and BALM-unpaired (*Fig 5a*). We also observed that ft-ESM learned the classification task much more quickly than base-ESM (*Fig 5d, 5e*), suggesting that the initial fine-tuning with paired sequences improved the models ability to adapt to the specificity classification task. Similar results were seen when using a synthetic dataset of HER2-specific and non-HER2-specific antibodies isolated from a heavy chain CDR3-swapped display library (*Table S1*). The second task, trained on a smaller dataset of approximately 2,000 paired antibodies (~1,000 in each class), involved binary classification of influenza (**Flu**)-specific and CoV-specific antibodies. In this task, BALM-paired is the best-performing model across all metrics except precision (*Fig 5b*). The ability of BALM-paired to outperform ft-ESM despite being a much smaller model is surprising and emphasizes the importance of focused pre-training with natively paired antibody sequences. The final specificity classification task, trained on approximately 3,000 antibodies (~1,000 per class), was a multi-class classification of Flu-specific, CoV-specific, and randomly selected healthy donor antibodies (*Fig 5c*). Similar to the second task, BALM-paired outperformed all other models across all metrics, followed closely by ft-ESM. Notably, BALM-unpaired outperformed base-ESM in this task, unlike previous tasks, but once again the models pre-trained with paired antibody sequences (BALM-paired and ft-ESM) each outperformed the equivalent model that was not pre-trained with paired antibodies (BALM-unpaired and base-ESM).

**DISCUSSION**
Current antibody language models are limited by their exclusive use of unpaired sequences and by inherent biases in publicly available antibody sequence datasets, which overrepresent certain disease states. The Jaffe dataset, with approximately $1.6 \times 10^6$ natively paired human antibody sequences from healthy donors, offers a unique opportunity to train an AbLM without these limitations. Given the relatively small size of this paired dataset, the benefits of training with natively paired sequences were not expected to overcome the shortage of data. Therefore, rather than attempting to train a state-of-the-art model using only natively paired data, we sought to determine how natively paired sequences could improve the training of AbLMs by training a matched set of models: BALM-paired and BALM-unpaired. In this controlled experiment, we show that natively paired training data substantially improves model performance and that these improvements are the result of BALM-paired learning immunologically relevant features that span both antibody chains.

Templated regions encoded by antibody germline segments were learned rapidly by BALM-paired, but the model struggled with untemplated regions including heavy chain CDR3s and regions with increased SHM. These results suggest that model training could be improved by incorporating more somatically mutated sequences and focusing training resources on untemplated regions. BALM-paired and BALM-unpaired both generate informative heavy chain embeddings that indicate their ability to learn antibody-specific features, grouping antibody embeddings primarily by mutation and secondarily by V gene use. In contrast, BALM-paired performs significantly better than BALM-unpaired on light chain embeddings. While the clustered light chain embeddings from BALM-unpaired do not segregate into well-formed clusters, those of BALM-paired are more similar to heavy chain embeddings, clustering primarily by mutation and secondarily by V gene. This suggests that BALM-paired is learning cross-chain features that improve light chain embeddings. The asymmetry with which cross-chain features influence model outputs, with light chain embeddings displaying much more obvious differences than heavy chains, is consistent with a growing body of evidence that the light chain partners of genetically similar heavy chains are themselves genetically similar [21]. This cross-chain information flow does not appear to be bilateral, however, as genetically similar light chains display "promiscuous" pairing with diverse heavy chains [27]. Thus, there is an immunological basis for the distinct improvement patterns observed with BALM-paired. We provide further evidence that BALM-paired is learning biologically relevant, cross-chain features by demonstrating markedly improved SHM prediction in one antibody chain when the natively paired chain, but not a germline-reverted variant of the natively paired chain, is provided as context. This implies a surprisingly sophisticated understanding of humoral immunity, learning that SHM-driven deviation from the germline template in one chain is a strong indicator of similar deviation in the paired chain.

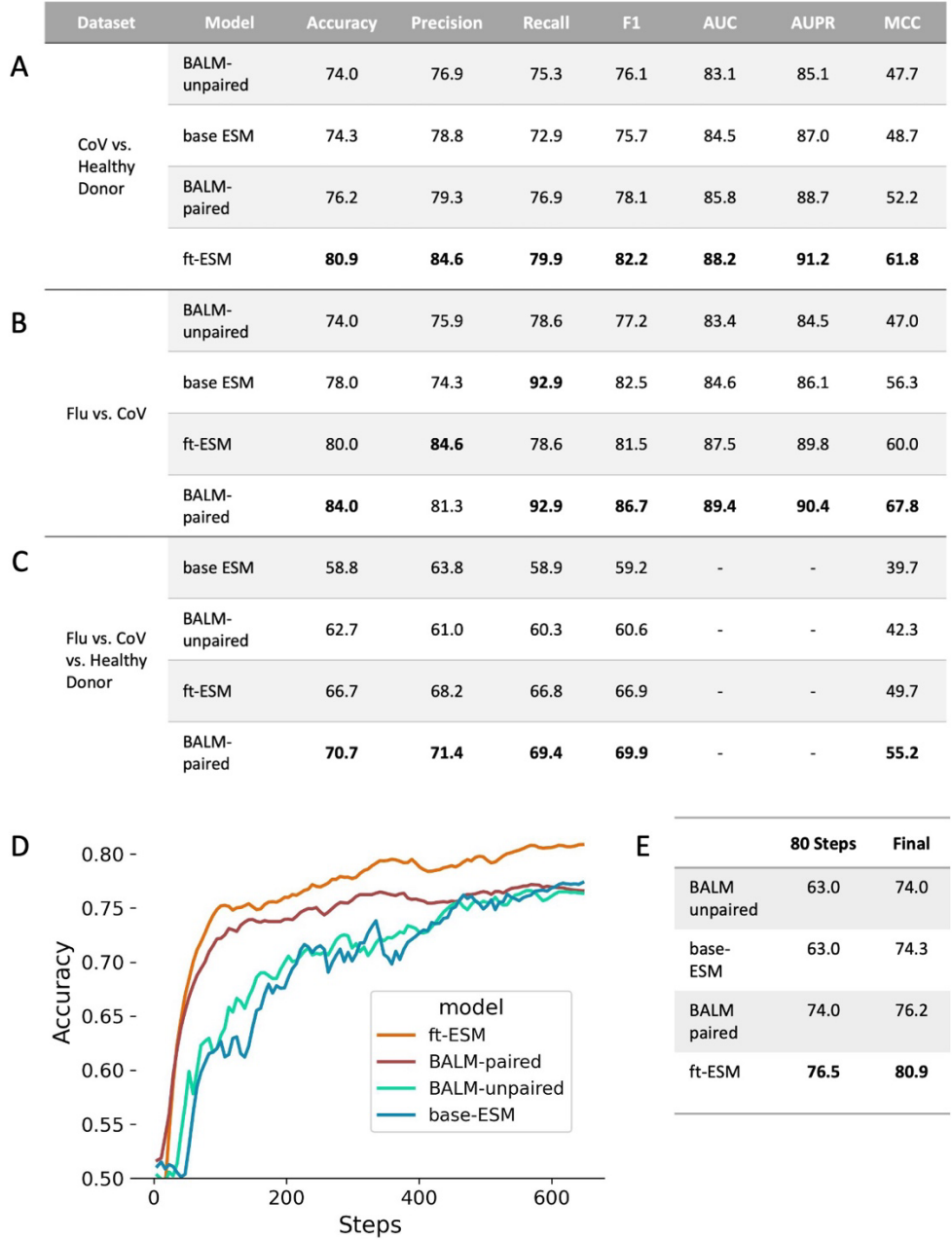

**Figure 5. Comparison of model performance on specificity classification tasks.** (A) Metrics of binary classification of CoV vs. healthy donor antibodies. (B) Metrics of binary classification of Flu vs CoV antibodies. (C) Metrics of multi-class classification of Flu vs CoV vs healthy donor antibodies. (D) Accuracy plotted against the number of sequences seen during training for the CoV vs. healthy donor classification. (E) Comparison of accuracy at 80 steps (~2500 sequences) versus the end of training for the CoV vs. healthy donor classification.

Although these results clearly demonstrate the benefits of training antibody language models with natively paired sequence data, in practice this is less straightforward, as the cost of generating paired antibody sequences is orders of magnitude higher than unpaired. To evaluate the feasibility of a middle-ground approach in which paired antibody sequences are used to supplement larger and more readily available training datasets, we fine-tuned the general protein language model ESM-2 [14] using the Jaffe dataset. The fine-tuned ESM-2 variant, but not the base ESM-2 model, focuses its cross-chain attention on immunologically important and structurally proximal regions of paired antibody sequences. This result is particularly encouraging, as it indicates that natively paired datasets – which, due to their high cost are necessarily limited in scale – can be supplemented with unpaired antibody sequences or general protein sequences while still allowing models to learn critically important cross-chain antibody features.

To demonstrate an example application of these paired models, we trained sequence classifiers to test the models' ability to perform three separate antibody specificity classification tasks. We observed that the paired models, BALM-paired and ft-ESM, consistently outperformed BALM-unpaired and base-ESM. It is noteworthy that BALM-paired (300M parameters) outperforms the larger ft-ESM (650M parameters) on complex multi-class classifications and when using smaller fine-tuning datasets. This suggests that pre-training exclusively with antibody datasets, or with mixed datasets more equally balanced between antibody and general protein sequences, may be a more promising strategy going forward than fine-tuning existing protein LMs.

Although the results from these binary and small multiclass classification tasks are impressive, it is not clear whether there is much practical use for models that can perform relatively simple "SARS-CoV-2 or not" classification tasks. Instead, the fact that these models can achieve such high accuracy means that there are learnable patterns of sequence-inherent properties that distinguish groups of antibodies with similar specificity. Herein lies what is likely the greatest value of these models: if we can better understand the feature patterns driving classification decisions, we have the opportunity to learn the fundamental immunological properties that define antibody specificity, with broad ramifications across infectious disease, autoimmunity, and cancer.

In summary, we report four important discoveries that will help guide the design and training of future state-of-the-art AbLMs. First, incorporating natively paired training data produces higher performing models by allowing models to learn cross-chain features. The native pairing of heavy and light chains is integral to the structure and function of each antibody and understanding features that span both chains is vital for performing downstream tasks such as antibody specificity classification. Second, AbLMs rapidly learn patterns associated with templated regions that are encoded by germline gene segments but struggle with untemplated regions. It is likely that training datasets enriched in somatically mutated sequences combined with antibody-specific training schemes that bias training resources toward untemplated regions such as CDR3s may directly address the most prominent model weaknesses. Third, mixed training datasets, which supplement paired antibody sequences with unpaired or general protein data, can help overcome the high cost and limited availability of natively paired datasets. Finally, LMs trained or fine-tuned using natively paired antibody sequences perform better on downstream classification tasks, suggesting a deeper and more generalizable understanding of human antibodies.

**METHODS**

*Training data.* For model pre-training, we used the largest publicly available dataset of natively paired human Ab sequences, comprising approximately $1.6 \times 10^6$ sequence pairs [21,22]. All paired Ab sequences in this dataset were recovered from circulating B cells from healthy adult human donors and were not selected or enriched for binding to any particular antigen. Raw sequences were annotated with abstar [28] and the amino acid sequence of each V(D)J region was extracted. Sequence pairs were filtered to remove duplicates and non-productive sequences, resulting in 1,335,854 filtered pairs. 90% of the filtered pairs were used for training, with 5% held out for evaluation and an additional 5% for testing.

For specificity classification training, three datasets were used. CoV antibody sequences were obtained from CoV-AbDab [29]. Flu antibody sequences were obtained from Wang et al [30], filtered for paired sequences only. Randomly selected antibodies from the memory B cell repertoire of healthy adult donors were obtained from the control dataset of Hurtado et al [31]. Amino acid sequences were clustered at 95% identity and cluster centroids were maintained. From here, these datasets were used to form three unique datasets to use for specificity classification tasks and labeled according to their antigen specificity: CoV vs Healthy Donor (total 18,092 sequences), CoV vs Flu (total 2,486 sequences), and CoV vs Flu vs Healthy Donor (total 3,729 sequences). Sequences were labeled according to their antigen specificity (or non-specificity, for the healthy donor sequences). All three datasets were randomly split, with 96% used for training, 2% held out for evaluation, and 2% held out for testing.

*BALM training.* We separately trained two BALM variants, BALM-paired and BALM-unpaired, using the HuggingFace transformers library [32]. Both models used a slightly modified version of the RoBERTa-large architecture [33], with 24 layers, 16 attention heads per layer, a hidden size of 1024, and an intermediate (feed-forward) size of 4096.

The vocabulary contained 25 tokens: one for each of the 20 amino acids and five special tokens: `<s>`, `</s>`, `<pad>`, `<unk>`, and `<mask>`. Inputs to BALM-paired were concatenated heavy and light chain sequences separated by a `</s>` token and padded to a maximum input length of 512. Inputs to BALM-unpaired were individual heavy or light chain sequences, padded to a maximum length of 256. None of the paired or unpaired sequences exceeded the maximum input length of their respective model, so truncation was not required. To equalize training steps between BALM-paired and BALM-unpaired, the total batch size of BALM-unpaired (512) was twice that of BALM-paired (256).

Both models were trained using a masked language model (**MLM**) objective. Briefly, when given an input for which some positions have been masked, the model is asked to predict the masked tokens based only on the context provided by the non-masked tokens. For each input, 15% of the tokens were uniformly selected for masking. Of the selected tokens, 80% were replaced with a `<mask>` token, 10% were replaced with a randomly selected amino acid token, and 10% were left unchanged. Masking was performed dynamically to avoid using the same mask across epochs [33]. BALM-paired and BALM-unpaired were each trained for 500,000 steps (approximately 100 epochs) on eight NVIDIA A100 GPUs, which equates to approximately 5 days per model. The peak learning rate was 4e-4, with a linear warmup over the first 30,000 steps and a linear decay thereafter.

*Analysis of model embeddings.* The output embedding of a model with input length *L*, hidden size *H*, and *N* input sequences, is a matrix of the shape *N* x *H* x *L*. For each BALM model, the dimensionality of the final layer output embedding was reduced by averaging over the *L* dimension as previously described [16,18], producing an *N* x *H* matrix. A uniform manifold approximation and projection (**UMAP**) embedding [25] was computed for the averaged embeddings matrix for each model in Python 3.9, using the umap-learn package [26]. UMAP plots were visualized in Python 3.9 using matplotlib. For BALM-paired, the subset of the output embedding matrix corresponding to either the heavy chain or light chain was extracted prior to averaging so that only the embeddings for the chain of interest were used to compute the UMAP. This ensures an "apples-to-apples" comparison between the embeddings of BALM-paired (for which the raw embeddings contain both heavy and light chains) and BALM-unpaired (for which the raw embeddings contain only a single chain).

*ft-ESM training.* We fine-tuned the pre-trained 650-million parameter ESM-2 model which is based on the RoBERTa architecture [33] and has 33 layers with 20 attention heads per layer [14]. The 650-million parameter model was chosen (rather than the larger, higher-performing 3B or 15B parameter ESM-2 variants) to reduce the likelihood of overfitting due to the small training dataset and allow for faster training despite memory constraints. Inputs were concatenated heavy and light chain sequences separated by two `<cls>` tokens and were tokenized with the standard ESM-2 vocabulary and padded to a maximum length of 320. None of the paired or unpaired sequences exceeded the maximum input length, so truncation was not required. The total batch size was 256. The model was trained using a MLM objective, as described above for BALM model training. The peak learning rate was 4e-4, with a linear warmup over the first 30,000 steps and a linear decay thereafter. The model was trained for 150,000 steps on eight NVIDIA A100 GPUs, which equates to approximately 7 days.

*Analysis of cross-chain attention.* The attention values of the model were extracted for each position of the input antibody sequence, across each head and layer of the model. The attention values were filtered to include only cross-attention (that is, position pairs for which the two positions are on different chains), and the retained attention values for each position pair were averaged across all 20 heads and 33 layers of the model. Based on this data, heatmaps were generated using seaborn [34] and matplotlib [35]. To map the cross-chain attention onto mAb structures, the total cross-chain attention was separately summed for each position in the heavy and light chain, resulting in a single attention vector per chain. These attention vectors were used to color residues by b-factor using PyMOL [36]. Attention step-plots were created using the summed attention vectors in Python using matplotlib [35].

*Specificity classification training.* BALM-unpaired, BALM-paired, ft-ESM, and base-ESM were fine-tuned with a sequence classification head for the downstream task of specificity prediction on two binary classifications (CoV vs Healthy Donor, CoV vs Flu) and one multi-class classification (CoV vs Flu vs Healthy Donor). For tokenization, models were tokenized with the standard tokenizer for the model type. BALM-paired and BALM-unpaired received concatenated heavy and light chain sequences separated by the `</s>` token while base ESM and ft-ESM were concatenated heavy and light chain sequences separated by two `<cls>` tokens. For BALM-unpaired only, two sequences longer than 253 tokens were removed since the maximum input size is 256 tokens. For the other three models, no truncation was necessary since all sequences were shorter than the models maximum input length. Models were trained for 1 epoch with a total batch size of 64 across eight NVIDIA A100 GPUs, with a learning rate of 5e-5 and a linear warmup ratio of 0.1.

Metrics used for evaluation of the binary classifications were accuracy, precision, recall, F1, Area Under the Receiver Operating Characteristic Curve (**AUC**), Area Under the Precision-Recall Curve (**AUPR**), and Matthews Correlation Coefficient (**MCC**). For the multi-class classifications, evaluation metrics were accuracy, macro-precision, macro-recall, macro-F1, and MCC. Plot of accuracy against model steps was based on wandb logging data and was smoothed with a weight of 0.75.

*Data and Code Availability.* Model weights for BALM-paired, BALM-unpaired, and ft-ESM and the datasets used for training are available on Zenodo [37] under the CC BY-SA 4.0 license. The code used for data processing, model training, and cross-chain attention plots is available on Github (github.com/brineylab/BALM-paper) under the MIT license.

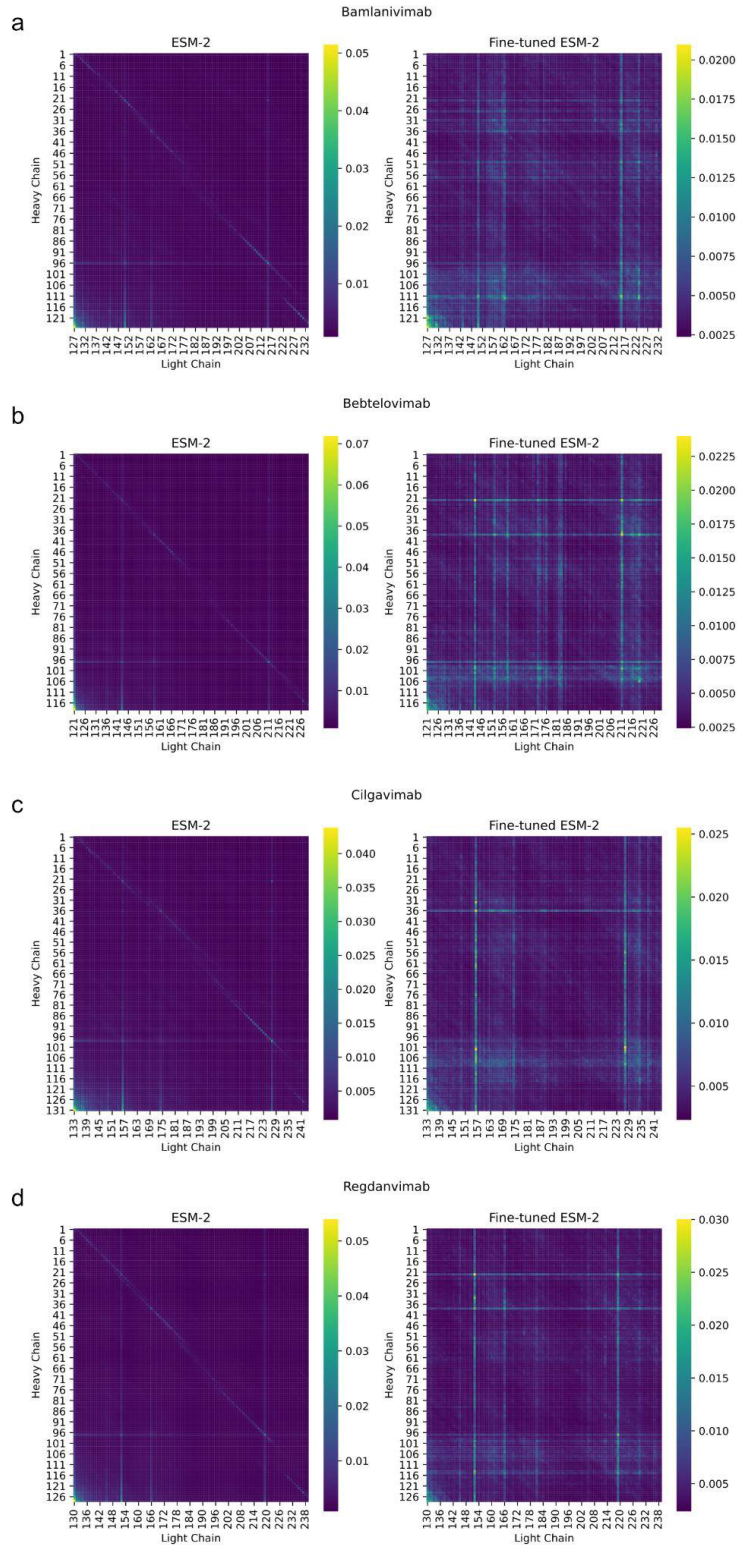

**Supplementary Figure 1. Cross-chain attention for selected therapeutic mAbs.** Four therapeutic mAbs against SARS-CoV-2 were processed using ESM-2 (right plots) or the fine-tuned ESM-2 variant: (a) Bamlanivimab, (b) Bebtelovimab, (c) Cilgavimab, and (d) Regdanvimab. Cross-chain attention was computed by averaging attention from all heads of each model layer.

| Dataset | Model | Accuracy | Precision | Recall | F1 | AUC | AUPR | MCC |
|---|---|---|---|---|---|---|---|---|
| HER2 vs Non-HER2 | BALM-unpaired | 73.8 | 67.5 | **95.7** | 79.2 | 89.2 | 89.3 | 51.8 |
| | base ESM | 82.5 | 83.7 | 82.5 | 83.1 | 89.8 | 90.3 | 64.9 |
| | BALM-paired | 82.9 | 84.7 | 82.0 | 83.3 | 90.2 | 91.0 | 65.8 |
| | ft-ESM | **84.5** | **85.5** | 84.6 | **85.1** | **91.5** | **91.9** | **68.9** |

**Supplementary Table 1. Comparison of model performance on specificity classification of HER2 vs non-HER2 binding antibodies.** The HER2 specificity classifier was trained on synthetic CDR3-swapped antibody sequence dataset from [38].